\pdfoutput=1
\documentclass[twocolumn,showpac,aps,prb,superscriptaddress,longbibliography]{revtex4-2}
 
\usepackage{dcolumn}

\usepackage[utf8]{inputenc} 
\usepackage{amsmath}
\usepackage{amssymb}
\usepackage{bm}
\usepackage{txfonts}
\usepackage{mathdots}
\usepackage{graphicx}
\usepackage{xcolor}
\usepackage[T1]{fontenc}
\usepackage{xcolor}
\usepackage{float}

\newcommand{\braket}[2]{\left \langle #1 \middle| #2 \right \rangle}

\DeclareUnicodeCharacter{2212}{-}
\DeclareUnicodeCharacter{2009}{\,}
\DeclareUnicodeCharacter{3B4}{$\delta$}
 
\begin{document}


 
\title{Intercalation-induced states at the Fermi level and the coupling of intercalated magnetic ions to conducting layers in Ni$_{1/3}$NbS$_2$}


\author{Yuki Utsumi Boucher}
\email{yutsumi@ifs.hr}
\affiliation{Institute of Physics, Bijenička c. 46, 10000 Zagreb, Croatia}

\author{Izabela Bia\l o}
\altaffiliation[Current affiliation: ]{Physik-Institut, Universität Zürich, Winterthurerstrasse 190, CH-8057 Zürich, Switzerland}
\affiliation{AGH University of Krakow, Faculty of Physics and Applied Computer Science, 30-059 Krakow, Poland}

\author{Mateusz A. Gala}
\affiliation{AGH University of Krakow, Faculty of Physics and Applied Computer Science, 30-059 Krakow, Poland}

\author{Wojciech Tabiś}
\affiliation{AGH University of Krakow, Faculty of Physics and Applied Computer Science, 30-059 Krakow, Poland}

\author{Marcin Rosmus}
\affiliation{Solaris National Synchrotron Radiation Centre, Jagiellonian University, Czerwone Maki 98, 30-392 Krakow, Poland}
\affiliation{Faculty of Physics, Astronomy, and Applied Computer Science, Jagiellonian University, Łojasiewicza 11, 30-348 Krakow, Poland}

\author{Natalia Olszowska}
\affiliation{Solaris National Synchrotron Radiation Centre, Jagiellonian University, Czerwone Maki 98, 30-392 Krakow, Poland}

\author{Jacek J. Kolodziej}
\affiliation{Solaris National Synchrotron Radiation Centre, Jagiellonian University, Czerwone Maki 98, 30-392 Krakow, Poland}
\affiliation{Faculty of Physics, Astronomy, and Applied Computer Science, Jagiellonian University, Łojasiewicza 11, 30-348 Krakow, Poland}

\author{Bruno Gudac}
\affiliation{Department of Physics, Faculty of Science, University of Zagreb, Bijenička c. 32, 10000 Zagreb, Croatia}

\author{Mario Novak}
\affiliation{Department of Physics, Faculty of Science, University of Zagreb, Bijenička c. 32, 10000 Zagreb, Croatia}

\author{Naveen Kumar Chogondahalli Muniraju}
\affiliation{Institute of Physics, Bijenička c. 46, 10000 Zagreb, Croatia}
\affiliation{Institute of Nuclear Physics PAN, Radzikowskiego 152, 31-342 Kraków, Poland}

\author{Ivo Batistić}
\affiliation{Department of Physics, Faculty of Science, University of Zagreb, Bijenička c. 32, 10000 Zagreb, Croatia}

\author{Neven Barišić}
\affiliation{Department of Physics, Faculty of Science, University of Zagreb, Bijenička c. 32, 10000 Zagreb, Croatia}
\affiliation{Institute of Solid State Physics, TU Wien, 1040 Vienna, Austria}

\author{Petar Popčević}
\email{ppopcevic@ifs.hr}
\affiliation{Institute of Physics, Bijenička c. 46, 10000 Zagreb, Croatia}

\author{Eduard Tutiš}
\email{etutis@ifs.hr}
\affiliation{Institute of Physics, Bijenička c. 46, 10000 Zagreb, Croatia}

\begin{abstract} 
The diversity of magnetic orders that appear in layered magnetic materials is of great interest from the fundamental point of view and for applications. In particular, the magnetic sublayers, introduced by intercalation into van der Waals gaps of the host transition-metal dichalcogenide (TMD), are known to produce various magnetic states, with some being tunable by pressure and doping. The magnetic sublayers and their magnetic ordering strongly modify the electronic coupling between layers of the host compound. Understanding the roots of this variability, starting from the underlying electronic structure, is a significant challenge. Here we employ the angle-resolved photoelectron spectroscopy at various photon energies, the {\it ab initio} electronic structure calculations, and modeling to address the particular case of Ni-intercalate, Ni$_{1/3}$NbS$_2$. We find that the bands around the Fermi level bear the signature of a strong yet unusual hybridization between NbS$_2$ conduction band states and the Ni 3$d$ orbitals. The hybridization between metallic NbS$_2$ layers is almost entirely suppressed in the central part of the Brillouin zone, including the part of the Fermi surface around the $\mathrm{\Gamma}$ point. Simultaneously, it gets very pronounced towards the zone edges. It is shown that this behavior is the consequence of the rather exceptional, {\it symmetry imposed}, spatially strongly varying, {\it zero total} hybridization between relevant Ni magnetic orbitals and the neighboring Nb orbitals that constitute the metallic bands. We also report the presence of the so-called $\beta$-feature, discovered only recently in two other magnetic intercalates with very different magnetic orderings. In Ni$_{1/3}$NbS$_2$, the feature shows only at particular photon energies, indicating its bulk origin. Common to prior observations, it appears as a series of very shallow electron pockets at the Fermi level, positioned along the edge of the Brillouin zone. Unforeseen by {\it ab initio} electronic calculations, and its origin still unresolved, the feature appears to be a robust consequence of the intercalation of 2H-NbS$_2$ with magnetic ions.
  
\end{abstract}


\maketitle


\section{Introduction}

  For decades, the interaction between localized magnetic moments and itinerant electronic subsystems has been central to condensed matter research. 
  In particular, it has been regularly and extensively discussed in the context of heavy fermions \cite{Steglich1979} and high-temperature superconducting cuprates, whose discovery has posed many puzzles to the research community \cite{Bednorz1986, Keimer2015, Kamihara2006, DeLaCruz2008, Barisic2022}. 
  Even earlier, various materials, where the same interaction may be expected, have been discovered and cataloged after inquiries appropriate for the times. 
  The magnetic intercalates of transition metal dichalcogenides (TMDs) are one such family of compounds \cite{vandenBerg1968, Anzenhofer1970, Friend1977}.
  Probably the best known and most numerous of those are based on 2H-NbS$_2$, with formula \textit{M}$_{1/3}$NbS$_2$ (\textit{M}: 3\textit{d} transition metal). 
  These compounds are particularly attractive due to the high quality of the crystals achieved at the 1/3 concentration of intercalated atoms and the range of exciting intertwined magnetic and transport properties they exhibit \cite{Friend1977, Parkin1980a, Parkin1980b, Parkin1983}. 
  In recent times, along with the development of new experimental techniques and theoretical concepts, there has been a revived interest in the 2H-NbS$_2$ intercalates. 
  With the interaction mentioned above taking the limelight, the question of the electronic properties of these materials seems far from being settled. 
   Apart from the common questions regarding the charge transfer between metallic layers of the host compound and the intercalated species, and the resulting spin and charge states of the latter, the other questions remain to be addressed: How do the magnetic ions connect the otherwise weakly connected metallic layers, and how their magnetic states and fluctuations affect the electronic structure in the vicinity of the Fermi level. Some of these questions are to be addressed here. 

  The host compound 2H-NbS$_2$ is a superconductor below 6 K \cite{VanMaaren1966}. 
  The intercalation of transition metal ions between NbS$_2$ layers suppresses the superconductivity and often results in complex magnetic orderings. 
  In \textit{M}$_{1/3}$NbS$_{2\ }$the transition metal ions occupy the octahedral voids between NbS$_2$ layers and form a $\mathrm{\sqrt{3}}\mathrm{\times}$$\mathrm{\sqrt{3}}$ superlattice rotated by 30$\mathrm{{}^\circ}$ in respect to the 1$\mathrm{\times}$1 unit cell of 2H-NbS$_2$. 
  Intercalation of the late transition metals Fe, Co, and Ni leads to predominant antiferromagnetism, whereas the early transition metals V, Mn, and Cr tend to order ferromagnetically \cite{Friend1977, Parkin1980a, Parkin1980b, Parkin1983}. 
  In some cases, the type of long-range magnetic order is found tunable by small changes in the concentration of the intercalate \cite{Wu2022}.
  Recent studies have revealed the chiral helimagnetic structure in Cr$_{1/3}$NbS$_2$ and Mn$_{1/3}$NbS$_2$ \cite{Togawa2012, Dai2019}, whereas weak ferromagnetism along the \textit{c}-axis was found to accompany the dominant antiferromagnetic ordering in Co$_{1/3}$NbS$_2$ \cite{Ghimire2018}. 
  Furthermore, a large anomalous Hall effect has been reported in the latter compound, the origin of which is still debated \cite{Ghimire2018, Smejkal2020, Tenasini2020, Mangelsen2021, Tanaka2022}.

  The angle-resolved photoelectron spectroscopy (ARPES) is expected to help significantly in understanding the electronic properties of these crystals. 
  The current study uses ARPES to address the case of Ni$_{1/3}$NbS$_2$. 
  Along the way, we will contrast this compound to Co$_{1/3}$NbS$_2$, recently studied by several groups \cite{Tanaka2022, Yang2022, Popcevic2022}. 
  At first, going from Co-intercalated to Ni-intercalated material can appear equivalent to doping, imposing a shift of the Fermi level within the almost-rigid electronic band structure of the host compound.
  A deeper comparison finds more significant and intricate differences, and the difference in band fillings in the two compounds is not one of them.
  Notably, the antiferromagnetic orderings in Ni$_{1/3}$NbS$_2$ and Co$_{1/3}$NbS$_2$ differ significantly. The magnetic structure of Co$_{1/3}$NbS$_2$ was solved previously \cite{Parkin1983} and confirmed recently \cite{Tenasini2020}. 
  The reported antiferromagnetic ``hexagonal ordering of the first kind'' breaks the 120${}^\circ \ $rotational symmetry of the crystal and doubles the unit cell. 
  The only attempt of refining the magnetic structure in Ni$_{1/3}$NbS$_2$ resulted in the claim that no additional magnetic peaks are observed in the antiferromagnetically ordered phase \cite{vanLaar1971}. 
  Such a finding is in accordance with the theoretically proposed magnetic structure \cite{Polesya2019} where Ni magnetic moments are ferromagnetically arranged within Ni-layers, whereas their direction alternates between subsequent Ni-layers. 
 In that way, the magnetic unit cell stays the same as the crystallographic one, as the crystallographic unit cell already comprises two Ni layers. 
  The same magnetic structure will be used in our calculations of the electronic structure of Ni$_{1/3}$NbS$_2$, to be presented further below. 
   Also, in contrast to Co$_{1/3}$NbS$_2$, no evidence of the out-of-plane ferromagnetic moment nor anomalous Hall effect has been reported for Ni$_{1/3}$NbS$_2$. 
  Other differences and similarities should be sought in the electronic structure, which we explore here. 
  We will mainly focus on the states near the Fermi level, dominantly influencing the low-energy electronic process. 
   Motivated by the recent finding of a new "$\beta$-feature" in the vicinity of the Fermi level, around the $\mathrm{K}$-point of the first Brillouin zone in Cr$_{1/3}$NbS$_2$ and Co$_{1/3}$NbS$_2$ \cite{Sirica2016, Sirica2020, Tanaka2022, Yang2022, Popcevic2022, Qin2022}, we also explore its possible existence in Ni$_{1/3}$NbS$_2$.       
  Experimentally detected, the feature is unforeseen by any electronic structure calculation, whereas the arguments in favor of its origin in strong electronic correlations have been presented recently \cite{Popcevic2022}. 
   Despite the mechanism of its appearance not being fully understood, the $\beta$-feature has been observed only in compounds with intercalated magnetic ions, with the orbitals of the intercalated magnetic ions participating in the phenomenon \cite{Sirica2016, Sirica2020, Popcevic2022}.

 In this work, we devote special attention to the influence of intercalated ions on the Nb 4\textit{d} conduction bands. 
  Recently a large amplification of the interlayer hybridization has been found in Co$_{1/3}$NbS$_2$ and Cr$_{1/3}$NbS$_2$ \cite{Popcevic2022, Qin2022}, effectively changing the character of these bands near the $\mathrm{\Gamma }$ point from quasi-two to three-dimensional. 
   However, it will turn out that the coupling between NbS$_2$ layers and the intercalated magnetic ions works very differently in Co$_{1/3}$NbS$_2$ and Ni$_{1/3}$NbS$_2$.   


\section{Experiment and Methods}

  Single crystals of Ni$_{1/3}$NbS$_2$ were grown by chemical vapor transport method using iodine as a transport agent \cite{Friend1977}. 
  The grown single-crystals were characterized by electrical resistivity measurement, x-ray diffraction, and magnetic susceptibility. 
    Electrical resistivity and magnetic susceptibility measurements revealed the antiferromagnetic ordering temperature, $T_N=$95 K, higher than earlier reported 90 K \cite{Friend1977,Parkin1980a}.
  The residual resistivity ratio (R(300 K)/R(2 K)) of 19, relatively high for this class of materials \cite{Friend1977, Parkin1980b}, also indicates the high quality of crystals.
  The measured room temperature lattice parameters are \textit{a }= 5.7609 (5) Å, and \textit{c }= 11.9014 (7) Å and agree well with previously reported values \cite{Anzenhofer1970}. 
  Using the Laue diffraction, samples were oriented along the high symmetry $\mathrm{\Gamma}$M$_{0}$ direction (which corresponds to $\mathrm{\Gamma}$M direction of 2H-NbS$_2$ Brillouin zone) and glued to Ti flat plates.
  The pristine surfaces were obtained by \textit{in situ} cleaving. 
  A sample was placed within the preparation chamber, cleaved under ultra-high vacuum (better than 2$\mathrm{\times}$10${}^{-10}$ mbar), and immediately transferred to the measurement chamber with the base pressure of 8$\mathrm{\times}$10${}^{-11}$ mbar. 
  The experiment was performed at the UARPES (currently URANOS) beamline of the SOLARIS synchrotron, using the photons in the 34--82 eV energy range. 
  The measurements were conducted at 20 K, well below the magnetic ordering temperature.
  The photoelectrons were collected using the hemispherical analyzer (Scienta-Omicron DA30) and the multichannel plate coupled to a CCD detector. 
  The overall energy resolution was set to $\mathrm{\sim}$30 meV. Linearly polarized light was used in the experiment. 
  The electric field vector of the incident light was parallel to the incident plane defined by the light propagation vector and was normal to the sample surface. 
  The details of the experimental setup can be found in Ref. \onlinecite{Popcevic2022}. 

  The band structure for Ni$_{1/3}$NbS$_2$ was calculated using the Quantum ESPRESSO (QE) Density Functional Theory package \cite{Giannozzi2009, Giannozzi2017}. 
  We have used the kinetic energy cut-off of 160 Ry for wavefunctions and 800 Ry for charge density and potentials. 
  To some level, the electronic correlations on Ni ions were accounted for using the DFT+$U$ approach proposed by Cococcioni and Gironcoli \cite{Cococcioni2005}. 
  The Hubbard interaction, $U$ = 7.7 eV, was determined through first-principles calculation \cite{Timrov2018}. 
  The ultrasoft pseudopotentials from pslibrary-0.3.1 \cite{DalCorso2014} are based on Perdew-Burke-Ernzerhof exchange-correlation functional revised for densely packed solids \cite{Perdew2008, Perdew2009}.
  The k-mesh within the Brillouin zone was $10\times 10\times 5$, without shift, whereas the Fermi-energy discontinuity was smeared as proposed by Marzari-Vanderbilt \cite{Marzari1999} using the broadening of 0.01 Ry.


\section{40 eV ARPES and DFT spectra}


\begin{figure*}[t!] 
\includegraphics[width=\textwidth]{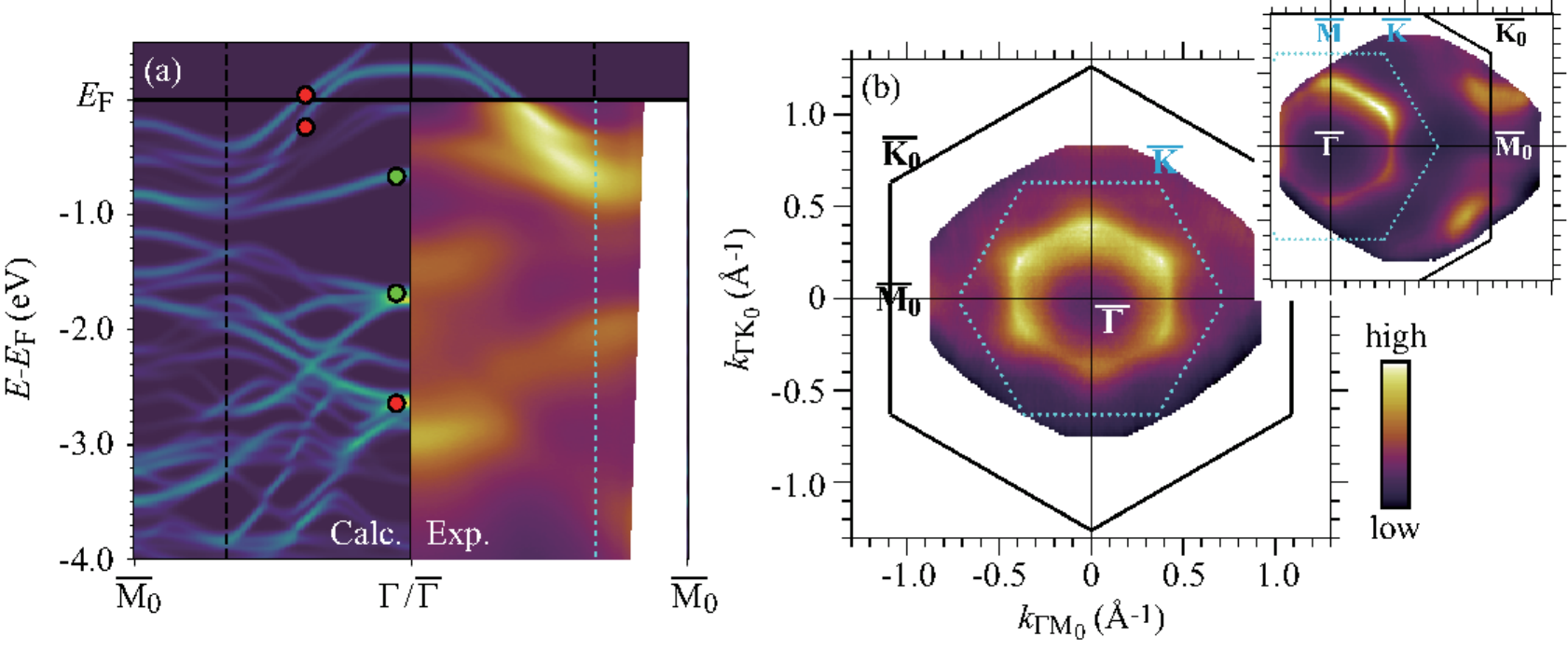}
\caption{(Color online) 
The ARPES signal in $k$-space. 
Here and throughout the paper, $\mathrm{\Gamma}$, K$_0$, and M$_0$ denote the usual characteristic points of the first Billiouin zone of the 2H-NbS$_2$, whereas K stands for the usual characteristic point of the three-fold smaller first Billiouin zone of Ni$_{1/3}$NbS$_2$. $\overline{\mathrm{\Gamma}}$,  ${\overline{\mathrm{M}}}_0$, ${\overline{\mathrm{K}}}_0$, and ${\overline{\mathrm{K}}}$ represent the corresponding points of the surface Brillouin zones. 
(a) ARPES intensity plot of Ni$_{1/3}$NbS$_2$ along $\overline{\mathrm{\Gamma}}{\overline{\mathrm{M}}}_0$ direction measured with h$\nu$ = 40 eV is shown on the right. 
  The measured spectra are compared to the unfolded DFT-calculated spectra for magnetically ordered Ni$_{1/3}$NbS$_2$ (shown on the left). 
  The vertical dashed lines represent the boundaries of the Ni$_{1/3}$NbS$_2$ surface Brillouin zone. 
  The red (light green) dots on the left panel mark the electronic states with predominant contributions from niobium (sulfur) orbitals. 
  (b) Fermi surface map of Ni$_{1/3}$NbS$_2$, also obtained by using h$\nu$ = 40 eV. 
  The black hexagon and the smaller blue hexagon shown in the dashed line represent the first Brillouin zone of 2H-NbS$_2$ and Ni$_{1/3}$NbS$_2$, respectively. The inset shows another scan that extends to the boundary of the (big) first Brillouin zone of 2H-NbS$_2$}
\label{fig1} 
\end{figure*}

  Figure 1 shows the ARPES intensity plot obtained using the incident photon energy (h$\nuup$) of 40 eV, along with the results of the performed density functional theory (DFT) calculations. 
  The right-hand side of Fig. 1 (a) presents the ARPES intensity plot of Ni$_{1/3}$NbS$_2$ measured along $\overline{\mathrm{\Gamma}}{\overline{\mathrm{M}}}_0$ direction.
 The notation regarding the $k$-space points is explained in the caption of Fig. 1. 
  In agreement with other ARPES results in the \textit{M}$_{1/3}$NbS$_2$ series \cite{Battaglia2007, Sirica2016, Tanaka2022, Yang2022, Popcevic2022}, no evidence of the band splitting can be observed at the boundary of Ni$_{1/3}$NbS$_2$ Brillouin zone (vertical dashed line in Fig. 1 (a)). 
  However, the part of the conduction band signal weakens significantly after crossing the dashed boundary. 

  The DFT-calculated spectra are shown on the left-hand side of Fig. 1 (a) \footnote{The conventional illustration of the calculated band structure of Ni$_{1/3}$NbS$_2$ can be found in Supplemental Material.}. 
   These spectra are shown as {\it unfolded} into the Brillouin zone of 2H-NbS$_2$ for a more straightforward comparison with the experimental data. 
   Two approaches to unfolding were used. In the most direct approach, we use the {\it unfold-x} procedure \cite{Popescu2012,Pacile2021,Bonfa} on the Quantum Espresso results. The second approach is based on the tight-binding (TB) parametrization of the DFT-calculated bands. 
  The TB parametrization, also used elsewhere in the paper, is based on the Wannier90 code \cite{Pizzi2020}.
  The two approaches produce the same results regarding the spectra in Fig. 1(a). 
   However, the TB approach may also be exploited to get insight into the orbital content of particular electronic states. 
   In that respect, the red and light green dots in Fig. 1(a) mark the energies of electronic states with the predominant contribution of niobium and sulfur orbitals, respectively. 
  The two red dots closest to the Fermi level mark the Nb 4$d$ bands. These signals resemble the anti-bonding and the bonding Nb 4$d$ bands of 2H-NbS$_{2\ }$ studied previously, both experimentally and through DFT calculations \cite{Sirica2016, Youbi2021, Heil2017, Popcevic2022}.

The band fillings are notoriously hard to assess precisely from ARPES scans.  Even under the assumption of Luttinger's theorem being applicable in a particular system \cite{Luttinger1960},  the need arises to perform ARPES at the dense mesh of $k$-points in three dimensions instead of addressing only the high symmetry cross-sections.  Including the several high-symmetry cross-sections, comparable to $\mathrm{\Gamma}-\mathrm{M}_0$ cross-section in Fig. 1(a) helps only slightly \cite{ Popcevic2022}, and the experimental-only means to assess the bands-filling may require extensive soft X-ray ARPES scans, providing more precise $k_z$ mapping of the electronic spectra \cite{Strocov2003}.   
In intercalated systems, the observation of the full Fermi surface in the $k$-space is further obscured by a large number of Fermi surface segments, a weak signal of some bands within the first Brillouin zone of the intercalated material, coming from ARPES matrix elements. In addition, and not unlike our system, the strong electron correlations tend to add complications that go beyond a simple band structure.  
In practice, the best course in estimating the band filling in the intercalated system partly relies on comparing the ARPES spectra in the intercalated and the host compound and comparing measured and DFT-calculated spectra. In the intercalated systems, the comparison between the latter two is further complicated by the fact that the intercalation is expected to perturb the spectrum of the host material extending over the Brillouin zone, several times bigger than the one calculated for the intercalated system. Thus, there is a need for unfolding the calculated spectra to the original Brillouin zone of the host material before making the comparison.  
In addition, in our particular case, the problem with the precise determination of the Fermi surface volume arises from the $\beta$-feature appearing in the measured spectra, to be addressed in the following section, which lacks its DFT-calculated counterpart. 

   The DFT results are used to calculate the charge of the Ni ion  
by integrating the charge density around the Ni site, with the result of   +1.6 $e$.
    The numerical evaluation of the conduction bands filling from DFT results produces the result of 5/6. 
   Compared to the host compound 2H-NbS$_2$, the extra 2/3 electrons per NbS$_2$ formula unit is found in conduction bands. 
     Within the rigid band picture, this would require the charge transfer of precisely 2 electrons from Ni ion into NbS$_2$ layers. 
     Thus, the result for the conduction bands filling seems to oppose the result for the charge of the Ni ion of +1.6 $e$, mentioned above.
     The resolution of this small puzzle lies in the fact that the situation in Ni$_{1/3}$NbS$_2$ very much departs from the rigid band picture. 
     The strong hybridization between intercalated ions and the layers of the host compound invalidates the simple relation between conduction bands filling and the charge state of the intercalated ion. 
     Further details are provided in Appendix B.

  Going to higher binding energies, we first address the band that separates from the conduction bands approximately halfway between ~${\overline{\mathrm{M}}}_0$ and ~$\overline{\mathrm{\Gamma}}$ points. 
  In the experimental side of Fig. 1(a), this band reaches the $\mathrm{\Gamma}$ point at the {\it binding energy} ($E_B$) of 0.4 eV, 
 i.e., at energy $E=E_{\rm F}-E_B=-0.4$ eV, below the Fermi level ($E_{\rm F}$).  

  The band that shows similar behavior in the calculated spectrum reaches $\mathrm{\Gamma}$ point at $E_{B}\approx $ 0.65 eV (marked by the higher light-green dot in Fig. 1(a)). 
  This band is the sulfur 3$p_{z}$ band, which is particularly sensitive, especially around the $\mathrm{\Gamma}$ point, to minor changes in the $c$-axis lattice constant. 
  This high sensitivity has been documented in the recent numerical studies in 2H-NbS$_2$, and its intercalates \cite{Youbi2021, Popcevic2022}. 
  Thus, the 0.25 eV shift in energy at the $\mathrm{\Gamma}$ point should not prevent us from making the correspondence between the observed band and the calculated band, otherwise behaving similarly.
  A similar shift in size and direction can be observed for the sulfur 3$p$ signal around 1.6 eV binding energy, marked by the lower light-green dot in Fig. 1(a). 
  Finally, the lowest red dot in Fig. 1(a) indicates the Nb-prevailing state around 3 eV at  $\mathrm{\Gamma}$ point and marks the only region in the lower part of the spectrum where the contribution from sulfur orbitals is not dominating. 

  Further similarities and differences between ARPES results for Ni$_{1/3}$NbS$_2$ and 2H-NbS$_2$ \cite{Sirica2016, Youbi2021} can be discussed in some detail, as it was done in the case of Co$_{1/3}$NbS$_2$ \cite{Popcevic2022}. 
  Instead, we focus on several essential features the Ni intercalation brings to the electronic spectrum. 

  Fig. 1 (b) shows the Fermi surface of Ni$_{1/3}$NbS$_{2\ }$measured using $h\nu$ = 40 eV. 
  The large black hexagon and the small blue hexagon in the dashed line represent the surface Brillouin zones of 2H-NbS$_2$ and Ni$_{1/3}$NbS$_2$, respectively. 
  As in the previous ARPES studies of Ni$_{1/3}$NbS$_2$ \cite{Battaglia2007, Tanaka2022}, the part of the Fermi surface centered at the $\mathrm{\Gamma}$ point is present in the form of a hexagonally shaped hole pocket. 
  Fig. 1 (b) does not extend wide enough to make visible the part of the Fermi surface centered around ${\overline{K}}_0$ points at the edge of 
the Brillouin zone of 2H-NbS$_2$.
  This part of the Fermi surface, evolving from the similar sections in 2H-NbS$_2$, is presented in the inset as a separate scan. 
  On the other hand, keeping in mind our interest for possible non-NbS$_2$$_{\ }$features appearing around the $\overline{K}$-point, at the edge of the Ni$_{1/3}$NbS$_2$ Brillouin zone, we note that no such signal is found in Fig. 1 (b). 
  However, as pointed out earlier \cite{Popcevic2022}, the search is to be continued by scans at other photon energies. This is precisely what we do next, by measuring the ARPES spectra in the 34 to 82 eV photon energy range in a few eV steps.


\begin{figure*}[t!] 
\includegraphics[width=\textwidth]{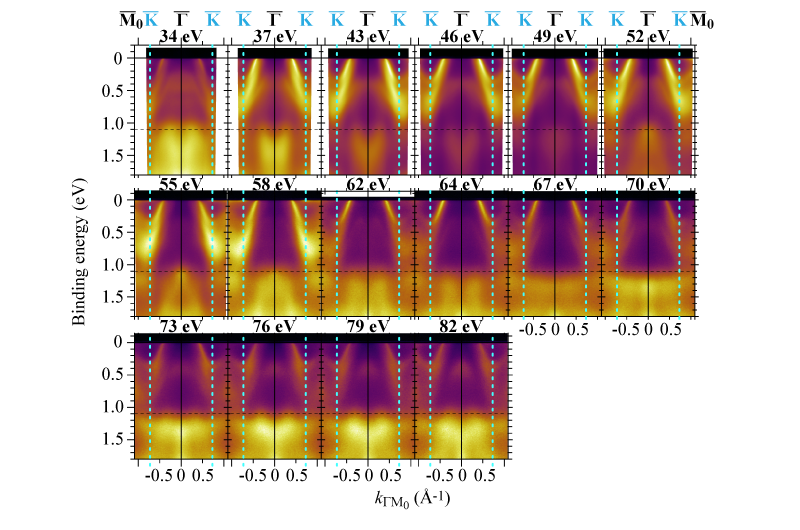}
\caption{(Color online) ARPES intensity plots of Ni$_{1/3}$NbS$_2$ measured along $\overline{\mathrm{\Gamma}}{\overline{\mathrm{M}}}_0$ direction using various incident photon energies from 34 to 82 eV. 
  The vertical dashed lines represent the boundaries of the Ni$_{1/3}$NbS$_2$ Brillouin zone. The total intensity of each ARPES spectrum, extending over a measured span of electronic momenta and the binding energy between 0 and 4 eV, is subsequently normalized to the same value.
  The signal very narrow in energy around the Fermi level, showing at the left and the right edges of the 58 eV panel, is the first trace of the $\beta$-feature, to be followed in other figures.     
}
\label{fig2} 
\end{figure*}


\section{The $\beta$-feature in the nickel intercalate}

  Fig. 2 shows ARPES intensity plots of Ni$_{1/3}$NbS$_2$ measured along $\overline{\mathrm{\Gamma}}{\overline{\mathrm{M}}}_0$ line using various photon energies. 
  What is immediately apparent are the changes between the panels that occur in the S 3$p$ bands located between $E_B=1$ and 1.8 eV. 
  These changes indicate a strong $k_z$ dispersion of these bands. 
  In contrast, the Fermi level crossings of the Nb 4$d$ bands are almost insensitive to the variation of the photon energy, pointing at the two-dimensional character of the Fermi surface, as expected for very anisotropic host material 2H-NbS$_2$ \cite{Youbi2021}. 
  
   The more important feature to be spotted in these spectra is the appearance of the shallow electron pocket that shows at the Fermi level at the boundary of Ni$_{1/3}$NbS$_2$ Brillouin zone (dashed vertical line), especially evident in $h\nu $=58 eV panel.
   These shallow electron pockets resemble those observed in Co$_{1/3}$NbS$_2$ \cite{Popcevic2022, Tanaka2022, Yang2022}, following the Brillouin zone boundary of the intercalate, and exceptionally well pronounced around its $\overline{K}$-point corners. 
  Apart from Co$_{1/3}$NbS$_2$, a similar feature has been observed also in Cr$_{1/3}$NbS$_{2\ }$\cite{Sirica2016, Sirica2020, Qin2022}. The signal/pocket observed around the $\overline{K}$-point was given the name ``$\beta$-feature'' \cite{Sirica2020, Popcevic2022}, which we also adopt here.


\begin{figure*}[t!] 
\includegraphics[width=\textwidth]{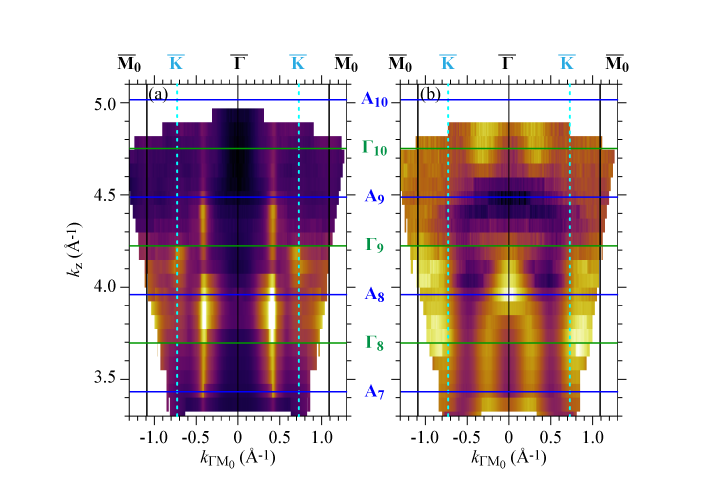}
\caption{(Color online) ARPES intensity plots along $k_z$ direction at (a) the Fermi level and (b) 1.1 eV below the Fermi level along the cuts indicated by the black dashed lines in Fig. 2. 
  The vertical dashed lines represent the boundary of the Brillouin zone of Ni$_{1/3}$NbS$_2$. 
  The horizontal green lines and labels ($\mathrm{\Gamma} _{n+1}$) mark the $k_z$ levels corresponding to  $\mathrm{\Gamma}_{n+1}$ $\equiv$ (0, 0, 2$\pi/c \times n$).
The horizontal blue lines and labels $\mathrm{A_{n+1}}$ mark the $k_z$ levels corresponding to $\mathrm{A}_{n+1} \equiv (0, 0, \pi/ c +2 \pi/ c \times  n$).
}
\label{fig3} 
\end{figure*}

  It is somewhat easier to identify the $\beta$-feature in Fig. 3, which presents ARPES intensity plots at constant energy along the \textit{k}$_{z}$ ($c$-axis) direction. 
  The same data presented in Fig. 2 are now used to plot the \textit{k}$_{z}$ band dispersion shown in Fig. 3, with the $k_z$ values obtained from the photon energy using the usual formula, $k_z=\sqrt{ (E_k{{\mathrm{cos}}^2 \theta +V_0)\ }2m /{{\hslash }^2}}$. 
  Here \textit{$E_{k}$}, \textit{$\theta$}, \textit{m}, and \textit{V}$_{0}$ denote the kinetic energy of a photoelectron, the angle between the surface normal and photoelectron's wave-vector, the free electron mass, and the inner potential, respectively \cite{Himpsel1983}. 
  The inner potential was set to \textit{V}$_{0}$ = 14 eV
, previously used in cases of Cr$_{1/3}$NbS$_2$ and Co$_{1/3}$NbS$_2$ \cite{Sirica2016, Popcevic2022}. It is worth to mention that $V_0$ might vary among different intercalates. The variation between ionic charges of magnetic ions is prone to strongly affect the electrostatic potential energy drop over the dipolar layer at the crystal surface. The versions of Fig.3 done for the inner potential $V_0$ range from 14 eV to 19 eV are provided in the Supplemental material
. The figures do not differ very much, with the differences being mostly perceived as shifts in $k_z$.  

  In the Fermi level scan in Fig. 3 (a), the $\beta $-feature appears close to the Ni$_{1/3}$NbS$_2$ Brillouin zone boundary, marked by dashed vertical lines at $k_{\parallel }$\textit{ }$_{\ }$= $\mathrm{\pm}$0.727 Å${}^{-1}$ $(\overline{\mathrm{K}}$-point). 
  This signal can be followed within the \textit{k}$_{z}$$_{\ }$= 4.0-4.2 Å${}^{-1}$ range around ${\mathrm{\Gamma }}_9$. 
  Apart from this $k_z$ dependence, the quasi-2D character of the Fermi surface originating from ``ordinary'' Nb 4\textit{d} bands is evident from Fig. 3 (a). 
  The signals corresponding to Nb 4\textit{d} bands cross the Fermi level at $k_{\parallel }$$_{\ }$= $\mathrm{\pm}$0.4 Å${}^{-1}$${}^{\ }$ throughout the measured $k_z$ range. 

  Fig. 3 (b) shows the constant energy surface cut at $E_B=1.1$ eV. 
  An oval-shaped constant energy surface elongated along the $k_z$ direction can be clearly distinguished with a pronounced $k_z$ dispersion. 
   The signal probably combines contributions of more than one sulfur $3p$ band. 
  A more direct insight may require an experiment covering a broader $k_z$ momentum range, using the incident photon energy extending into the soft X-ray region.

  Motivated by findings related to Fig. 3 (a) and to further examine the $\beta$-feature, we performed another Fermi-level scan using a photon energy of 58 eV. 
  The result of the scan is shown in Fig. 4. (a). 
  Along with the hexagonal hole pocket centered at the $\overline{\mathrm{\Gamma}}$ point, the $\beta $\textit{-}feature shows as smaller triangular electron pockets at each $\overline{\mathrm{K}\ }$corner of the small Ni$_{1/3}$NbS$_2$ Brillouin zone. 
  The shallowness of these pockets can be appreciated upon inspecting Figs. 4 (b) and (c). 
  These panels show the ARPES intensity profiles along $\overline{\mathrm{K}}$-$\overline{\mathrm{M}}$-$\overline{\mathrm{K}}$ and ${\overline{\mathrm{M}}}_0$-$\overline{\mathrm{\Gamma }}$-${\overline{\mathrm{M}}}_0$ directions, indicated by light green arrows in Fig. 4 (a).

  The feature showing around $\overline{\mathrm{K}\ }$points in Fig. 4 (a) is somewhat weaker and less structured than the one recently found in Co$_{1/3}$NbS$_2$ \cite{Tanaka2022, Yang2022, Popcevic2022}. 
  Nevertheless, it shares several qualities with the features observed recently in Co$_{1/3}$NbS$_2$ and Cr$_{1/3}$NbS$_2$  \cite{Sirica2016, Sirica2020, Tanaka2022, Yang2022, Popcevic2022, Qin2022}.  
  Centered around the same points in the $(k_x, k_y)$ plane, the $\beta$-feature has always been found as a relatively weak signal, confined around the Fermi level and occupying a very narrow energy window relative to other bands.    
  These similarities appear despite the three materials having different magnetic ordering types and significant electronic structure differences, some of which are to be addressed further below.  
  Our finding indeed supports the possible universality of the $\beta$-feature in magnetic intercalates of TMDs, as suggested in Ref. \onlinecite{Popcevic2022}.
  As elaborated there, this universality may stem from strong electronic correlations produced by the coupling between magnetic ions and the metallic subsystem. 
  The strong electron correlations are expected to arise primarily because the low energy excitations in magnetic sublayers tend to dynamically reconfigure the charge and spin exchange between metallic subsystems \cite{Popcevic2023}. In turn, and on a slower scale, the itinerant electrons can affect the state of the magnetic subsystem, thus introducing the "long-term memory" effects for the itinerant subsystem. 
  It should be noted, however, that the recent ARPES study of the \textit{M}$_{1/3}$NbS$_2$ (\textit{M}= Fe, Co, Ni) series, using $h\nu$ = 120 eV, reveals the $\beta$-feature around $\overline{\mathrm{K}}$-point in Co$_{1/3}$NbS$_2$ but not in Fe$_{1/3}$NbS$_2$ and Ni$_{1/3}$NbS$_2$ \cite{Tanaka2022}.
  Part of the reason for the latter discrepancy from our results may lie in the values of the incident photon energy used. 
  The earlier study in Co$_{1/3}$NbS$_2$ \cite{Popcevic2022} and here reported dependence of the signal intensity on photon energy speak in favor of such a possibility. 

    In principle, the dependence of ARPES spectra on photon energy offers information on band dispersion along the \textit{k$_{z}$} direction. In Co$_{1/3}$NbS$_2$ the \textit{k$_{z}$} band dispersion has been reported for the Co-derived band near the Fermi level \cite{Yang2022, Tanaka2022}. 
  The \textit{k$_{z}$} dependence of the $\beta$-feature may be considered a possible reason for it being seen only within a particular photon energy range. 
  However, other factors can also affect the ARPES signal, such as the photon energy-dependent transition matrix element influenced by the symmetry of the underlying electronic wavefunctions. 
   The $k_z$ range covered in our experiment by varying the photon energy exceeds the size of a single Brillouin zone, as evident from Fig. 3, which includes several replicas of the $\mathrm{\Gamma}$ point along the $k_z$-axis direction. More precisely, the $9^{th}$ replica of the first Brillouin zone in $k_z$-direction extends between points $\mathrm{A}_8$  and $\mathrm{A}_9$ in Fig. 3, with the point $\mathrm{\Gamma}_9$ in its center. Likewise, the $8^{th}$  and $10^{th}$ replicas are centered around $\mathrm{\Gamma}_8$  and $\mathrm{\Gamma}_{10}$.   
  
   The $\beta$-feature in Fig. 3(a) appears around ${\mathrm{\Gamma }}_9$, but does not show around ${\mathrm{\Gamma}}_8$ and ${\mathrm{\Gamma}}_{10}$. 
   Accordingly, the mere $k_z$ dispersion is probably not the main reason for the $\beta$-feature showing in a particular photon energy range.
   Further ARPES experiments over an extended photon energy range, possibly using soft X-rays, may help to resolve the question entirely.


\begin{figure*}[t!] 
\includegraphics[width=\textwidth]{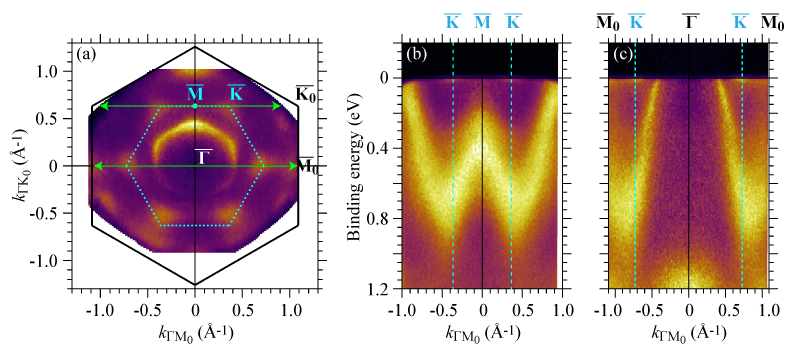}
\caption{(Color online) (a) Fermi surface map of Ni$_{1/3}$NbS$_2$ measured with h$\nu$ = 58 eV. 
  The pockets centered around $\overline{\mathrm{K}}$ and equivalent points in the Brillouin zone appear well pronounced here, witnessing the $\beta$-feature in Ni-intercalate.       
  The ARPES intensity plots along (b) $\overline{\mathrm{K}}$-$\overline{\mathrm{M}}$-$\overline{\mathrm{K}}$ and (c) ${\overline{M}}_0$-$\overline{\mathrm{\Gamma }}$-${\overline{\mathrm{M}}}_0$ direction, indicated by light green allows in panel (a).}
\label{fig4} 
\end{figure*}


\section{Qualitative differences in couplings between different magnetic ions and TMD layers}

  The second rather significant difference between the electronic structures of 2H-NbS$_2$ and Co$_{1/3}$NbS$_2$ is related to Nb 4\textit{d} conduction bands  \cite{Popcevic2022}. 
  A much larger splitting in energy between Nb 4\textit{d} bonding and anti-bonding bands was found in the central part of the Brillouin zone in Co$_{1/3}$NbS$_2$ than in 2H-NbS$_2$. 
  In 2H-NbS$_{2}$, {\it both} bands cross the Fermi level upon approaching the $\mathrm{\Gamma}$ point.    
  In Co$_{1/3}$NbS$_2$, only the anti-bonding band shows such crossing, while the bonding band is found submerged 0.3 eV below the Fermi level at the $\mathrm{\Gamma}$ point. 
  The latter was primarily attributed to strongly amplified hybridization between NbS$_2$ layers, with the intercalated Co ions serving as hybridization bridges \cite{Popcevic2022}. 
  The electron transfer from Co ions into NbS$_2$ layers was found to play a minor role in positioning the bonding band in Co$_{1/3}$NbS$_2$ below the Fermi level. No such splitting caused by intercalation-amplified hybridization between NbS$_2$ layers is visible in Ni$_{1/3}$NbS$_2$. No Nb 4\textit{d} bonding band is identified below the Fermi level at the $\mathrm{\Gamma}$ point. 
  The splitting between conduction bands in Ni$_{1/3}$NbS$_2$ is minimal at their crossing of the Fermi level. 
  On the other hand, the spotting between Nb-bands gets very pronounced upon approaching the zone boundary, as evidenced by Fig. 1(a). 

  The DFT calculations agree with such a finding: The calculated electronic spectra show strong intercalation-induced splitting between the bonding and the anti-bonding bands in Co$_{1/3}$NbS$_2$ \cite{Popcevic2022}, and very small splitting in the vicinity of the $\mathrm{\Gamma}$ point in Ni$_{1/3}$NbS$_2$ (see. Fig.1(a)). 
  This large difference between the two compounds is surprising since the Ni-Nb and Co-Nb hybridizations are expected to be similar in size, based on listed Ni and Co ionic radii \cite{Shannon1976} and the similar \textit{c}-axis lattice constants found in two compounds.
  The reasons for the discrepancy between the two systems can be explored by examining the Wannier90 tight-binding parametrization of the DFT-calculated spectra in terms of maximally localized orbitals and the hybridization integrals between them. 
   The Wannier90 parametrization in Ni$_{1/3}$NbS$_2$ produces the "maximally localized" Wannier functions that resemble the Ni 3$d$ and S 3$p$ atomic orbitals.
  As for the Nb-dominated conduction bands, one Wannier function per Nb site is produced. 
   These Wannier functions are dominantly composed of Nb 4$d$ atomic orbitals, with three neighboring Nb ions equally contributing to each Wannier function. 
   These effective three-legged, Y-shaped "Nb orbitals" are centered between three contributing Nb ions and extend toward their nuclei, as shown in Fig. 5(c). 
   For simplicity, in the remaining, we refer to them as "Nb Wannier orbitals", "Nb($Y$) orbitals", or just as the "Nb orbitals".  
   Note that the conduction bands in any 2H-NbS$_2$-based systems are dominantly composed of such Nb($Y$) orbitals.

   The participation of Ni ions in building the conduction bands comes through two minority-spin 3$d$ $e_g$ orbitals, $d_1$ and $d_2$. 
  These two orbitals are degenerate in energy, with their energy levels positioned less than 1 eV above the energy levels of the Nb orbitals and the Fermi level. 
  To a large degree, $d_1$ and $d_2$ resemble the standard 3$d_{z^2_1}$ and 3$d_{x^2_1- y^2_1}$ orbitals, with the coordinate system $(x_1,y_1,z_1)$ set by the ideal octahedron with the sulfur atoms at its vertices.       
  In essence, we find that the effective hybridization between any of the orbitals $d_1$ and $d_2$ and the Nb orbitals is comparable in size to analogous hybridization integrals found in Co$_{1/3}$NbS$_2$ \cite{Popcevic2022}.
  However, contrary to the Co system, the effective hybridization integrals between a Ni orbital and the Nb orbitals within a given Nb-plane vary in size and sign, as illustrated in Fig. 5. 
    Panels (a) and (b) in Fig. 5 show the hybridization integrals ${t_{\mathrm{Ni}(d_{1,2})-\mathrm{Nb(\mathit{Y})}_j}}$ between the Ni $3d$ $e_g$ orbitals $d_{1}$ and $d_{2}$, and the Nb($Y$) orbitals at position $j$ in the Nb layer {\it below} the Ni ion. 
    The sizes and signs of the hybridization integrals are shown through the areas of the circular surfaces and the colors of the disks centered at the positions of the Nb($Y$) orbitals. 
   Only the hybridization integrals for the three nearest-neighboring and the three next-to-nearest-neighboring Nb orbitals are shown. 
   The hybridization integrals to further Nb orbitals are very small in comparison. 
   It may be noted that the sign of the hybridization integral depends on the position of the Nb orbital. 
   Also, the absolute values of hybridization integrals for next-to-nearest-neighbors tend to be much bigger than for the nearest neighbors.
   Most importantly, it turns out that the {\it sum} of the hybridization integrals between any of the two Ni $3d$ $e_g$ orbitals and all the Nb Wannier orbitals Nb($Y$)$_j$ of a particular Nb plane, 
$\sum_j{t_{\mathrm{Ni}(d_{1,2})-\mathrm{Nb(\mathit{Y})}_j}}$,
is numerically indistinguishable from zero. 
   This vanishingly small sum is not accidental.   
   Instead, it is the consequence of the exact symmetry present in the system. 
  The claim can be exactly proved by constructing two linear combinations of two wavefunctions, $d_{o1,o2}\propto d_{1}\pm i d_{2}$, with $d_{o1}$ being shown in Fig. 5(c).
   By construction (see also Appendix C), these orbitals conform to the crystal symmetry, contrary to the orbitals $d_{1}$ and $d_{2}$. 
   Upon their rotation by $2\pi/3$ around the $c$-axis of the crystal, only their complex phase changes by $\pm 2\pi/3$.
   The corresponding sets of hybridization integrals possess the same symmetry: 
   The three hybridization integrals to Nb orbitals connected by the $\pm 2\pi/3$ rotation around the $c$-axis passing through the Ni-site have the same absolute value. 
   Only the complex phase of each hybridization integral is shifted by $\pm 2\pi/3$ relative to the other two. 
  Consequently, their partial sum is precisely zero. 
   The situation is illustrated in Fig. 5(d), where six hybridization integrals are positioned in the complex plane, and within the complex phase color wheel, forming two equilateral triangles.
   The symmetry argument appears in further detail in Appendix C.

  The vanishing sum of the hybridization integrals implies the zero value of the $(k_x,k_y)=0$ Fourier transform of the Ni-Nb hybridization. 
  Consequently, it also implies zero coupling between any of two Ni 3\textit{d} $e_g$ orbitals and the $(k_x,k_y)=0$ states of the Nb-layers. 
   The Fourier transform of the hybridization integral and the coupling to Nb states remains finite for non-zero $(k_x,k_y)$. 
  As a result, the splitting between the bonding and the anti-bonding bands produced by Ni intercalation \textit{vanishes} at the $\mathrm{\Gamma}$ point and remains small in the central part of the first Brillouin zone.
  It can be concluded that the striking difference in how Ni and Co ions hybridize/interact with the NbS$_2$ layers is mainly related to their spin and charge states acquired in similar local environments.
   The respective "starting/ideal" spin/charge states are Ni$^{+2}$ 3$d^8$ and Co$^{+2}$ 3$d^7$. 
   Different symmetries of hybridization configurations in Ni and Co compounds are set entirely by these "ideal" Ni/Co states.
   The substantial hybridization amplitude between Ni $3d$ $e_g$ orbitals and the conduction band Nb orbitals leads to Ni charge and spin state deviations from the ideal Ni$^{+2}$ 3$d^8$ state, but does affect the established symmetry. 
   These further deviations in charge are discussed in more detail in Appendix B.


\begin{figure*}[t!] 
\includegraphics[width=\textwidth]{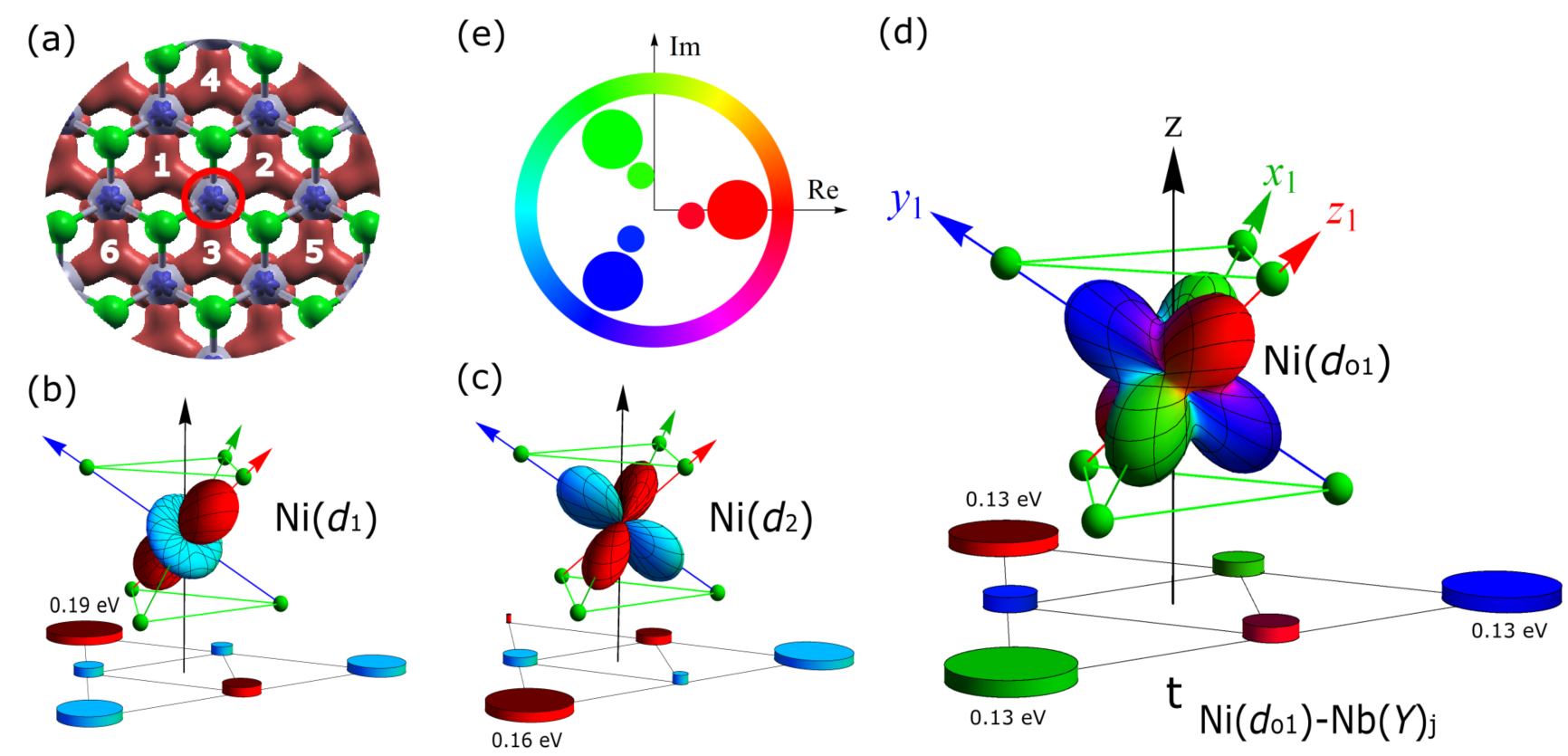}
\caption{(Color online) 
    (a) The crystal structure of 2H-NbS$_2$ as viewed along the $c$-axis direction. Grey balls represent the Nb atoms, and the sulfur atoms are shown in green. The three-legged Nb Wannier orbitals, each joining three neighboring Nb sites, are shown in dark-red (positive sign) and blue (negative sign). The red circle in the center marks the position of the magnetic Ni ion above the NbS$_2$ layer in Ni$_{1/3}$NbS$_2$.   
   (b), (c) The Ni $3d$ $e_g$ orbitals with energy above the Fermi level. The orbitals are shown in the usual way, with their absolute value being proportional to the distance of the surface point from the center of the Ni atom, and the sign is represented through color (red for +1, cyan for -1). The cylinders below each orbital represent the hybridization integrals between the orbital and the six Nb($Y$) Wannier orbitals, labeled by 1 to 6 in (a), in the Nb plane {\it below} the Ni ion. The cylinders are positioned at the centers of the Nb Wannier orbitals. The sizes and colors of the cylinders represent the sizes and signs of the hybridization integrals. The numbers next to some cylinders illustrate the absolute values of respective hybridization integrals in eV. Note the variation in absolute values and signs.  
   (d) The linear combination, $d_{o1}\sim d_{1}+id_{2}$, of the two $e_g$ orbitals. The complex phase of the wavefunction is coded in color. The complex phase color wheel is shown in (e).
   Contrary to $d_{1}$ and $d_{2}$,  $d_{o1}$ is the eigenfunction of the $2\pi/3$ crystal symmetry rotation. 
   The absolute values of hybridization integrals to Nb($Y$) orbitals at positions equivalent by crystal symmetry are the same in this representation, whereas their phases (the same color coding applies) differ by $\pm2\pi/3$. 
   (e) The same six cylinders (disks) are positioned within the complex phase color wheel, showing the equilateral triangular arrangement.
}
\label{fig5} 
\end{figure*}


\section{Conclusion}

  Using various photon energies, we have performed ARPES measurements in Ni$_{1/3}$NbS$_2$. 
  The calculated electronic spectra provide good guidelines for the analysis of our experimental findings.   
  The main elements of the observed band dispersion of Ni$_{1/3}$NbS$_2$ can be traced back to those of the host 2H-NbS$_2$. 
   The electron transfer from Ni atoms to NbS$_2$ layers is mainly responsible for the observed Fermi level shift within the 2H-NbS$_2$ conduction bands. 
   However, the 2H-NbS$_2$ rigid band picture cannot describe the electronic structure of Ni$_{1/3}$NbS$_2$.  
   Interestingly, the deviations from the rigid band picture qualitatively differ from those recently reported for Co$_{1/3}$NbS$_2$ \cite{Popcevic2022}. 
   In contrast to the cobalt case, the intercalation by nickel does not produce a strong splitting of the Nb 4$d$ conduction bands in the center of the Brillouin zone. However, we argue that both behaviors are consistent with strong hybridization between 3$d$ orbitals of the intercalated ion and Nb metallic states. We find that the hybridization between individual 3$d$ orbitals and neighboring Nb orbitals is comparable in size in both Ni and Co systems. What differentiates the two systems are the symmetries of the relevant 3$d$ states, the ones closest to the Fermi level. In the cobalt case the strong hybridization is reflected in the large conduction band splitting around the Brillouin zone center. Contrary to that, in the nickel case, the large conduction bands splitting caused by Ni orbitals occurs far from the center of the Brillouin zone but turns out exactly zero at the center of the Brillouin zone, and remains small around the zone center. The latter behavior seen in our ARPES data is understood through our DFT calculations and the symmetry analysis.    

   Regarding the newly discovered $\beta$-feature in Ni$_{1/3}$NbS$_2$, the pronounced differences between Ni$_{1/3}$NbS$_{2}$,  Co$_{1/3}$NbS$_2$, and Cr$_{1/3}$NbS$_2$ speak in favor of the anticipated universality of the $\beta$-feature \cite{Popcevic2022}.     
     Again, as in the case of Co intercalated compound, the $\beta $-feature is particularly pronounced around $\overline{\mathrm{K}}$-points of the Brillouin zone of the Ni intercalated material. 
  Nothing similar is expected from DFT calculations performed for  2H-NbS$_2$ or Ni$_{1/3}$NbS$_2$.  
   The $\beta$-feature signal's strength in Ni$_{1/3}$NbS$_{2}$ is found to depend strongly on the photon energy used in ARPES, pointing to the structure of its wavefunction along the $c$-axis, or its energy dispersion along the same direction. 
  The previously proposed \cite{Popcevic2022} role of strong electron correlations in creating the $\beta $-feature still remains to be confirmed. 


\section*{Acknowledgements}
   The Croatian Science Foundation has partly supported this work under the project numbers IP 2020-02-9666, UIP 2019-04-2154, and by the project Cryogenic Centre at the Institute of Physics -- KaCIF co-financed by the Croatian Government and the European Union through the European Regional Development Fund-Competitiveness and Cohesion Operational Programme (Grant No. KK.01.1.1.02.0012). 
  The research has been performed at the National Synchrotron Radiation Centre SOLARIS at the UARPES beamline (proposal No. 201046). 
  This publication was partially developed under the provision of the Polish Ministry and Higher Education project Support for research and development with the use of research infra-structure of the National Synchrotron Radiation Centre SOLARIS under contract nr 1/SOL/2021/2.
  The work at the TU Wien was supported by the European Research Council (ERC Consolidator Grant No 725521). The work at the University of Zagreb was supported by the Croatian Science Foundation under the project IP 2018 01 8912 and CeNIKS project co-financed by the Croatian Government and the European Union through the European Regional Development Fund - Competitiveness and Cohesion Operational Programme (Grant No. KK.01.1.1.02.0013). 
  The work at AGH University was supported by the National Science Centre, Poland, Grant OPUS: 2021/41/B/ST3/03454; the Polish National Agency for Academic Exchange under “Polish Returns 2019” Programme No. PPN/PPO/2019/1/00014; and the “Excellence Initiative–Research University” program for AGH University of Krakow.
  I.B. acknowledges support from the Swiss Confederation through Government Excellence Scholarships. E.T. acknowledges the support of the Croatian Science Foundation Project IP-2022-10-9423.


\section* {Appendix A. Unfolding from the Wannier90 representation}
The superstructure introduced by intercalation leads to the small first Brillouin zone that corresponds to the supercell in real space. The DFT calculations can provide the energy eigenvalues $E_{n,k}$ for the Kohn-Sham band $n$  and any wave vector $k$ within the small first Brillouin zone.
The ARPES spectra, however, are not confined to the wave vectors of the small Brillouin zone, or any Brillouin zone for that matter, but address the whole reciprocal space. The DFT results being  "folded" into a small part of the reciprocal space,  makes the comparison with measured spectra very difficult. The first step in comparing the DFT-calculated results with the spectra measured in ARPES goes in the direction of projecting the wavefunctions $\psi_{n,k}$ onto plain waves $e^{i k' r}$ at arbitrary wavevector $k'$. 

The approach to "unfolding" that we use in the paper uses the Wannier90 parameterization of the DFT results \cite{Pizzi2020}, relying on a handful of maximally localized Wannier wavefunctions ("orbitals") $\phi_m$ in the unit cell. The orbitals are usually centered around particular atoms and often acquire the form of the textbook atomic orbitals. Within the Wannier90 approach, the Kohn-Sham wave function $\psi_{n,k}$ follows the tight binding form for the crystal with several atoms per unit cell.
\begin{equation}
\psi_{n,k} (r)=\frac{1}{\sqrt{N} }\sum_{R,m} e^{i k R} \alpha_{n,k}^{(m)} \phi_m(r-R-d_m).
\label{eq:psink}
\end{equation}
Here $R$ stands for the vector of the Bravais lattice of the crystal, running over the positions of its unit cells. $N$ stands for the number of unit cells in the large-yet-finite crystal. The Wannier90 approach also provides the energy levels of the orbitals, $E_m$ as well as the overlap integrals $t_{m,m'}(R)$ between orbitals $m$ and $m'$ within the same unit cell ($R=0$), or within the unit cells separated by $R$. Within the usual scheme of the tight-binding approximation, these parameters lead to the energy bands $E_{n,k}$ that reproduce the DFT- calculated spectra very well,  and provide the orbital amplitudes $\alpha_{n,k}^{(m)}$ that enter the wave function $\psi_{n,k}$ in Eq.(\ref{eq:psink}). 
The projection of  the wave function $\psi_{n,k}$ on the plane wave  $\braket{r}{k'} = \frac{1}{\sqrt{V}} e^{I k' r} $, is easy to calculate, yielding 
\begin{equation}
\braket{k'}{\psi_{n,k}}=\sum_{m,G} \delta_{k',k+G} \alpha_{n,k}^{(m)}g_{m,k'} e^{ik'd_m}
\end{equation}
where $g_{m,k'}$ stands for the projection of the local orbital on the plane wave, 
\begin{equation}
g_{m,k'}\equiv\frac{1}{\sqrt{v_1}} \int dr e^{ik'r} \phi_m(r).
\label{eq:gm}
\end{equation}
and $G$ is the vector of the crystal's reciprocal lattice and $v_1$ denotes the volume of its unit cell. The absolute value of the projection determines the intensity of the unfolded spectrum at point $(k',E=E_{n,k})$,
\begin{equation}
w(k',E)=\sum_{n,k} \left|\braket{k'}{\psi_{n,k}}\right|^2 \delta(E_{n,k}-E),
\label{eq:wkE0}
\end{equation}
or, 
\begin{equation}
w(k',E)=\sum_{n,k} \left|\sum{m} \alpha_{n,k}^{(m)}g_{m,k'} e^{ik'd_m}\right|^2 \delta(E_{n,k}-E).
\label{eq:wkE}
\end{equation}
Here $k'$ stands for whatever vector in the reciprocal space, whereas $k=k'-G$ stands for its counterpart in the first Brillouin zone of the crystal. In order to make the variation of the weights visible in the image plot, one needs to replace the function $\delta(E)$ by the test function $\gamma(E,\eta)$ of a finite width $\eta$. In this paper we choose to use the Gaussian function for $\gamma(E,\eta)$ ,  $\gamma(E,\eta)= \frac{1}{{\gamma \sqrt {2\pi }}}e^{- E^2 /2\gamma}$, and $\eta=0.03$ eV.    
The advantage of the unfolding being done through the Wannier90 parametrization of the spectra, allows us to get additional insights by making the various choices for $g_m(m,k')$, instead of the one prescribed by Eq.(\ref{eq:gm}).
First, it should be remarked that the spectrum $w(k',E)$ as defined in Eq.(\ref{eq:wkE}) is not expected to simulate the ARPES spectra properly. In the simplest three-step approximation to ARPES \cite{Moser2017}, the ARPES matrix elements include the polarization wave vector $\epsilon$ of the photon beam, leading to 
\begin{equation}
g_{m,k',\epsilon}\equiv \frac{1}{\sqrt{v_1}} \int dr e^{ik'r} (\epsilon\cdot r) \phi_m(r)
\end{equation}
to appear in Eq.(\ref{eq:wkE}) instead of $g_{m,k'}$ \cite{Moser2017}. Second, the effect of the crystal surface is to be taken into account, leading to further modification of the spectrum \cite{Strocov2003, Moser2017, Strocov2012}.   
In fact, the simplest way to compare the DFT-calculated and ARPES spectra is to show all calculated bands in their unfolded variant, without some bands getting suppressed relative to others due to the particular symmetry or the size of the orbitals involved, or the photon polarization. This choice is the closest one to the usual plots of bands dispersion in the extended zone in the $k$-space, however with weights of particular points being modulated by the variations of atomic positions, orbital energy levels $E_n$ and the orbital overlap integrals $t_{m,m'}(R)$ being modulated by the superstructure. This choice is made in preparing the DFT-unfolded spectra in Fig. 1 (a). It corresponds to taking $ g_{m,k'}=1$ in Eq.(\ref{eq:wkE}).
Another way to profit from a particular choice of $g_m$'s, is to explore the contributions of particular orbitals/atoms by choosing $g_m=0$ for all the other orbitals. We use this possibility to focus on the contribution of the Nb orbitals to bands around the Fermi level. In Fig. 6 we show and compare the DFT+unfolded spectra in Ni$_{1/3}$NbS$_2$ and Co$_{1/3}$NbS$_2$ featuring only the contributions from Nb orbitals. Notably, contrary to the Co$_{1/3}$NbS$_2$ case, the Nb contribution close below the Fermi level is not observed around the $\Gamma$ point in Ni$_{1/3}$NbS$_2$.  Another way to plot the contribution of various orbitals to the unfolded spectra is provided in Supplemental Material, Fig. S2 (b).  


\begin{figure}[H]
\includegraphics[width=0.7\columnwidth]{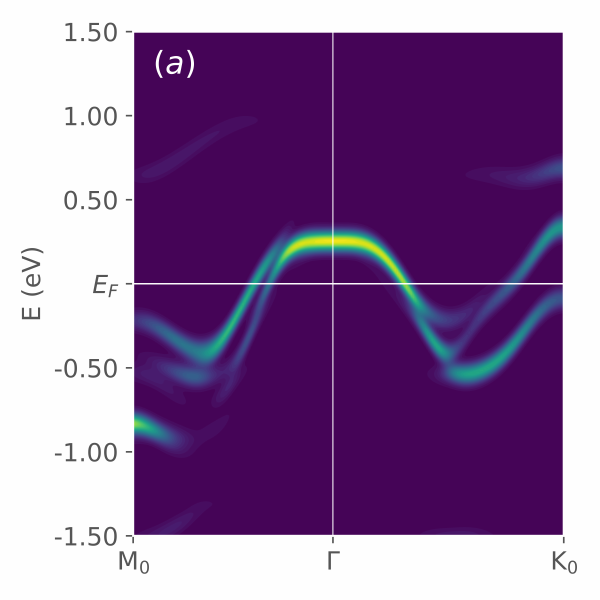}
\includegraphics[width=0.7\columnwidth]{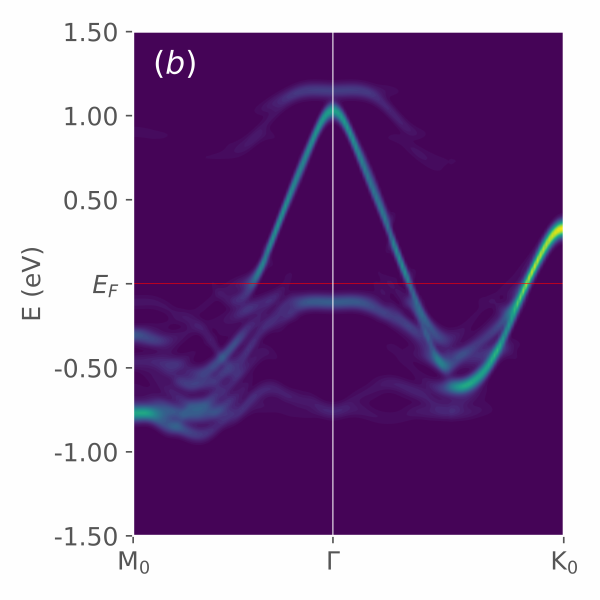}
\caption{(Color online) 
The unfolded spectra of the DFT-calculated results for Ni (a) and Co (b) intercalated systems, featuring only the contribution of Nb atoms. The great difference between the two figures is the result of very different ways in which Ni and Co ions connect the neighboring Nb layers, discussed in the main text. In particular, the big separation in energy between bands is visible at the $\Gamma$ point in the Co-based system, whereas almost no separation is found in the Ni-based system. The qualitative differences are also noticeable throughout the (large)  Brillouin zone of 2H-NbS$_2$. 
}
\label{fig:figNbOnly}
\end{figure}


\section*{Appendix B. Bands filling}

   Here, we discuss the effects of the intercalation on Ni charge and the conduction bands filling. 
   The direct approach, based on DFT-calculated "projected density of states" (PDOS), or the approach starting from the Wannier90 parametrization of the DFT spectra, may be used, giving the same results. 

\begin{figure}[H] 
\includegraphics[width=\columnwidth]{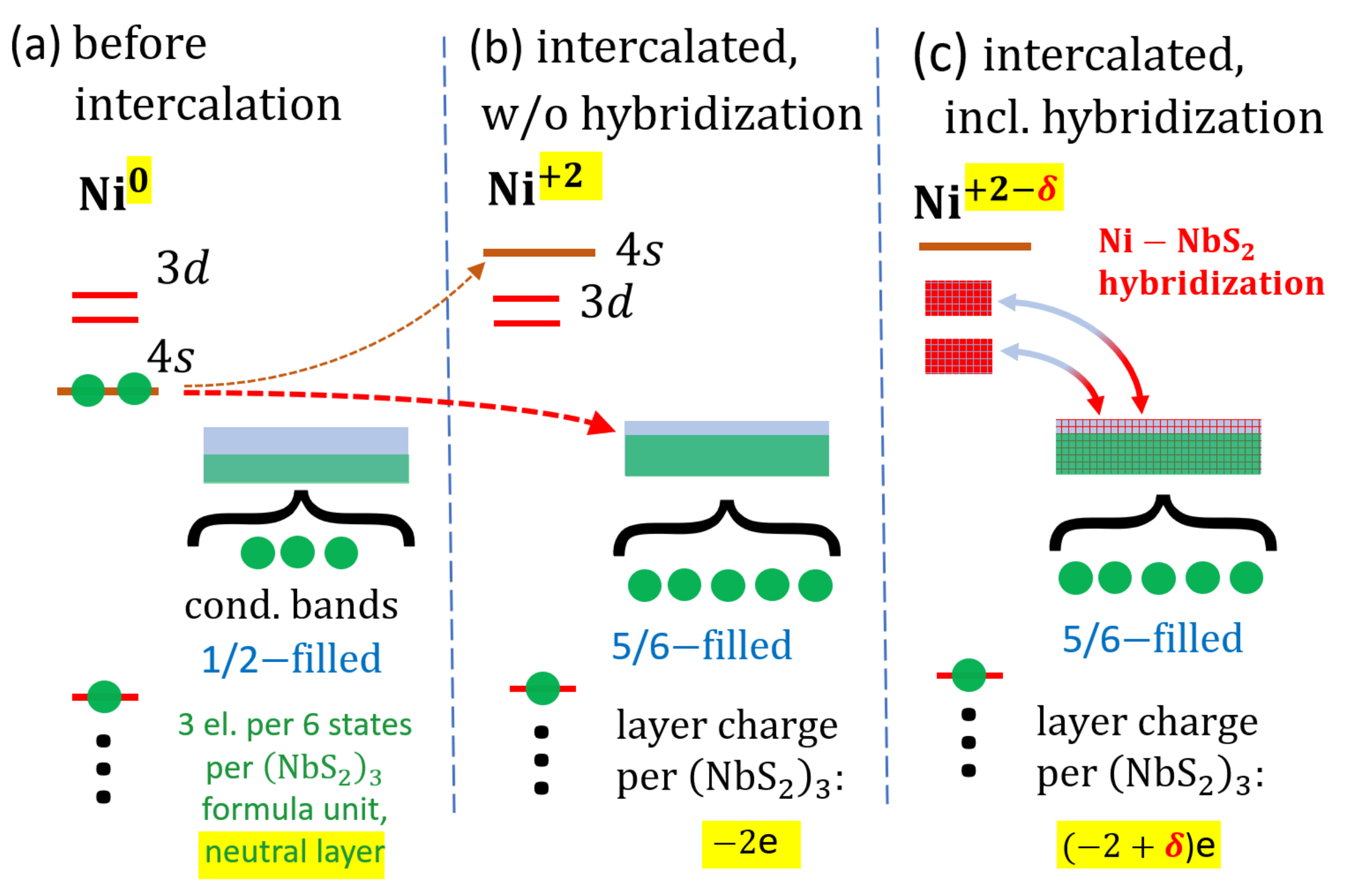}
\caption{(Color online)
The schematic view of Ni and NbS$_2$ subsystems. The Ni 3$d$ orbital levels are marked in red. The electrons and the filled portion in conduction bands are painted green. The NbS$_2$ empty states are shown in light blue. (a) The standalone magnetic Ni atom and NbS$_2$ crystal. (b) The transfer of 2 electrons from one Ni atom per 3 NbS$_2$ formula units, filling the conduction band to 5/6. (c) The hybridization mixes the Ni orbitals and the NbS$_2$ conduction bands states. In turn, the NbS$_2$ orbitals participate in the Ni-dominated bands. The number of states in the conduction bands and the number of electrons in those bands do not change. The band filling remains 5/6. The charge of the Ni ion profits from participating in occupied conduction band states by receiving extra $\delta$ electrons per Ni atom. The charge of the Ni ion is $+(2-\delta)$. 
}
\label{fig6} 
\end{figure}

  The integration of the density of states (DOS) obtained from the Quantum Espresso DFT calculation provides the value of the conduction bands filling factor of precisely 5/6. 
  This result is evident in the rigid-band picture, assuming that each Ni atom donates $q=2$ electrons into NbS$_2$ conduction band, resulting in the Ni$^{+2}$ ion. 
  For the $c_i=1/3$ concentration of intercalated atoms per NbS$_2$ formula unit, the conduction bands filling changes from $f_0=1/2$ to  $f=f_0 +  q\times c_i \times {1/2} =5/6$. The factor $1/2$ appears as the consequence of the spin degeneracy of band states.  
   In our case, the two electrons to be transferred to NbS$_2$ conduction bands reside in Ni 4$s$ orbitals of the standalone magnetic Ni atoms. 
  The transfer occurs upon positioning these atoms at the centers of sulfur octahedral voids. 
  Simultaneously, the lowest Ni orbitals above the Fermi level become the local minority-spin Ni $3d$ $e_g$ orbitals, $d_1\approx d_{z_1^2}$ and $d_2\approx d_{x_1^2-y_1^2}$. 
   The labels correspond to the local coordinate systems set by the sulfur octahedron (see also Fig. 5 and Appendix B).
   The final ingredient to account for is the hybridization between the intercalated atom's orbitals and the orbitals of the host material. 
   It may not be evident that hybridization invalidates the above relation between the charge of the intercalated ion and the conduction band filling. 
   The hybridization does not affect the band filling, as the number of states in conduction bands {\it and} the number of electrons hosted by those bands remain unchanged. 
   Conversely, the charge of the intercalated atom {\it does change} as its orbitals participate in the {\it filled} conduction bands states. 
   Our DFT calculation gives the value around 0.37 for the occupancy per Ni ion of its 3$d_{1 }$ and 3$d_{2}$ orbitals. 
  This calculation is done by integrating the partial density of states, derived by projecting the conduction bands wavefunctions to these atomic states (the L\"owdin method \cite{Lowdin1950}). 
  As already stated in the main text, the standard QE calculation of ionic charges, performed by integrating the spatial charge density distribution within the non-overlapping spheres around atomic nuclei, provides a compatible value of $ +1.6 (\approx 2-0.37)$ $e$ for the Ni ionic charge 
\footnote{A more elaborate discussion of the charge allocation to atoms within Quantum Espresso, using the L{\"o}wdin method, would also address the charge density and wavefunctions further away from the Ni nucleus. The related calculations tend to attribute some charge to Ni 4$s$ and Ni 4$p$ orbitals, with energy levels further above the Fermi level, through their hybridization with the states/bands deep below the conduction bands, presumably sulfur-dominated. Similarly, the hybridization of 3$d_{x_1^2-y_1^2}$ and 3$d_{z_1^2}$ orbitals with the states deep below the conduction band also provides some additional charge in these orbitals. We consider such details to be out of the scope of the present paper}.
  The message of this appendix is illustrated in Fig. \ref{fig6}.


\section*{Appendix C. The hybridization of Ni orbitals with the Nb layers}

 The Wannier90 parametrization of the DFT-calculated wavefunctions and electronic spectra in Co$_{1/3}$NbS$_2$ reveals only one relevant Co 3\textit{d} orbital close to the Fermi level that effectively (via intermediate sulfur orbitals) hybridizes with the Nb($Y$) orbitals \cite{Popcevic2022}.  
  This orbital is approximately described by the Co 3$d_{z^2}$ electronic state, oriented along the crystal's $c$-axis. The Co-Nb hybridization integral maximizes at approximately 0.24 eV for Nb($Y$)-orbitals nearest to the Co ion. 
  The total hybridization, $\sum_j{t_{\mathrm{Co}(d_{z^2})-\mathrm{Nb(\mathit{Y})}_j}}$, between the relevant Co orbital and all the Nb orbitals of the closest Nb layer amounts to approximately three times that value.
   In Ni$_{1/3}$NbS$_2$, the similar Wannier90 parametrization of the bands around the Fermi level produces two maximally-localized,  real-numbered Ni orbitals strongly hybridizing with the Nb($Y$) orbitals. 
   The orbitals $d_1$ and $d_2$ are the usual $e_g$ pair of states emerging in the octahedral environment provided by the sulfur ions. 
  As expected, they appear degenerate in energy for the ideal octahedron or the octahedron dilated along the $c$-axis, perpendicular to its two faces, as found in our crystal. 
  The resemblance between orbitals $d_1$ and $d_2$ and their counterparts within the ideal octahedron is very pronounced, 
$d_1 \approx d_{z_1^2}$ and 
$d_2 \approx d_{x_1^2 -y_1^2}$. 
  The coordinate system $(x_1,y_1,z_1)$ implied here points along the axes of the imaginary "ideal sulfur octahedron". 

  The calculated Ni($d_1$)-Nb($Y$)$_j$ and Ni($d_2$)-Nb($Y$)$_j$ hybridization integrals maximize around 0.2 eV in absolute value for specific Nb($Y$) orbitals. 
  However, for both Ni orbitals, the hybridization integrals appear with positive and negative signs. 
  Their sum, per single Ni orbital and single Nb layer, turns out numerically indistinguishable from zero within the numerical accuracy of the DFT+Wannier90 calculation.  
   The vanishing sum implies zero coupling between any of two Ni 3\textit{d} orbitals and the $(k_x, k_y)=0$ states of the Nb-layers.   
   Consequently, the splitting between bonding and anti-bonding bands produced by Ni intercalation vanishes at the $\mathrm{\Gamma}$ point and remains very small in the central part of the Brillouin zone. 

  It should also be remarked that the sums of \textit{absolute} values of hybridization integrals, per single Ni orbital and all Nb orbitals of the nearby layer, are only some 40\% smaller in Ni$_{1/3}$NbS$_2$ than in Co$_{1/3}$NbS$_2$. 
  Certain Ni-Nb hybridization integrals are indeed strongly pronounced, partly justifying the naive preliminary expectation regarding the comparable hybridization strengths in Co and Ni intercalates.

  The strictly vanishing sum of Ni-Nb hybridization integrals between relevant Ni orbitals and the metallic layer is not likely to be a mere numerical coincidence. 
  As we show now, it appears as the consequence of the symmetry that involves the $3d^8$ electronic configuration of the magnetic ion in the octahedral environment and the $2\pi/3$ rotation symmetry of the surrounding crystal. 

  The reasons for the vanishing average of the hybridization to nearby layers are not evident for Wannier90-supplied maximally localized orbitals. 
  It becomes more evident upon choosing the equivalent orbitals, $d_{o1}$ and $d_{o2} $, fully obeying the symmetry of the Ni ion environment, although slightly more extended in space.
  The crystal symmetry operator, $\mathcal{R}\mathrm{\equiv }\mathrm{Rot[}c,n\mathrm{\times }\mathrm{\ 2}\pi \mathrm{/}$3]$\mathrm{,\ }$ denoting the $2\pi \mathrm{/3}$ rotation around the $c$-axis of the crystal is essential in that respect. Indeed, the wavefunctions ${\mathcal{R}}^nd_{1}$   and  ${\mathcal{R}}^nd_{2}$ ($n$ denoting an integer) represent the choices equivalent to $d_{1}$ and $d_{2}\mathrm{\ }$for electronic structure description. 
  The same applies to the two linear combinations, 
\begin{align*}
 d_{o1}\propto\sum_{n=0}^2{e^{+in\frac{2\pi}{3}}\ \mathcal{R}^n\ d_1},\\
 d_{o2}\propto\sum_{n=0}^2{e^{-in\frac{2\pi}{3}}\ \mathcal{R}^n\ d_2},
\end{align*}
which are the eigenfunctions of the $\mathcal{R}$ operator by construction, and mutually orthogonal. 
  Upon normalization, and for the ideal octahedron, these two orbitals can be written as
\begin{align*}
d_{o1}=\frac{1}{\sqrt2}\left(d_{1}+id_{2}\right),\\
d_{o2}=\frac{1}{\sqrt2}\left(d_{1}-id_{2}\right). 
\end{align*}
  The choice of $d_{o1}$ and $d_{o2}$ for the Ni orbitals restores the crystal symmetry of the electronic structure parametrization, missing in the Wannier90 maximally localized approach.  

  Finally, we turn to $\mathrm{Ni}\left(d_{o1}\right)-\mathrm{Nb}(Y)_j$ and  $\mathrm{Ni}\left(d_{o2}\right)-\mathrm{Nb}(Y)_j$ hybridization integrals: 
  For a given Ni ion, these can be grouped into groups of three, each group containing Nb($Y$) orbitals connected by $\mathcal{R}$ rotations around the Ni ion. 
  Through crystal symmetry and our construction of $d_{o1}$ and $d_{o2}$, the related hybridization integrals are equal in magnitude and bear the phases proportional to unity, $e^{+i2\pi/3}$ and  $e^{-i2\pi/3}$. 
  Thus, the sum of hybridization integrals exactly vanishes for each group, rendering to zero the sum of the hybridization integrals between $d_{o1}$ ($d_{o2}$) orbital and any Nb layer. 
  However, as the hybridization varies from one Nb orbital position to another, the Fourier transform of the hybridization integral remains {\it finite} at finite $(k_x,k_y)$, being precisely zero only at $(k_x,k_y)=0$ point.     
    
  The illustration that involves the Ni ($d_{o1}$ orbital and six orbital positions in one Nb layer is provided in Fig. 5.
  Further illustrations can be found in the Supplemental Material. 

%

\end{document}